# Subdiffraction confinement and non-diffractive propagation of optical Stokes skyrmions enabled by a super-oscillatory metalens

**Jing He,[1#] Chengda Song,[1#] Wei Li,[2] Fangwen Sun,[3,4] and Guanghui Yuan[1,*]**

[1]Department of Optics and Optical Engineering, School of Physical Sciences, University of Science and Technology of China, Hefei, Anhui 230026, China

[2]Center for Micro and Nanoscale Research and Fabrication, University of Science and Technology of China, Hefei, Anhui 230026, China

[3]Chinese Academy of Sciences Center for Excellence in Quantum Information and Quantum Physics, University of Science and Technology of China, Hefei, Anhui 230026, China

[4]Hefei National Laboratory, University of Science and Technology of China, Hefei, Anhui 230088, China

[#]Authors with equal contributions

[*]ghyuan@ustc.edu.cn

**Abstract:**

Optical Stokes skyrmions have garnered extensive interest due to their intrinsic topological robustness and potential in informatics. However, most research remains confined to paraxial, low-numerical-aperture (low-NA) regimes, where their large transverse dimensions restrict broader applications. Under high-NA focusing, the polarization texture typically degrades or transforms abruptly as the beam traverses the focal region, hindering topology-preserving transport. In this work, we propose a strategy to generate a skyrmion needle field that maintains both subdiffraction confinement and non-diffractive propagation under high-NA conditions, thus preserving their topological characteristsics. Leveraging the polarization invariance of conventional optical needles, we realize the Stokes skyrmion needle using a single plasmonic metalens—designed to function as both a polarization filter and a super-resolving focusing element. Experimental and simulation results verify non-diffractive propagation over an extended depth of focus (up to $5\lambda$), while the Stokes-vector texture retained at subdiffraction scales throughout propagation. This skyrmion needle not only addresses previous propagation constraints but also opens new avenues for diffraction-unlimited information transport. Such skyrmion needles exhibit substantial potential in fields including light–matter interaction, optical metrology, and informatics.

## 1. Introduction

Topology plays a pivotal role in modern physics. Special topological structures—such as singularities,[1,2] knotted fields,[3,4] skyrmions,[5–7] and Möbius strips,[8,9]—have grown increasingly prevalent in research across diverse fields, including fluid dynamics,[10,11] magnetism,[12] optics,[13] and quantum physics.[14] Among these structures, skyrmions represent a class of topological entities that manifest as quasiparticles. First introduced in the 1960s to describe the topological structure of nucleons,[15] their topological properties are defined by an integer called the *skyrmion number*— a parameter that quantifies how many times a three-dimensional vector field in real space wraps around the unit sphere. In the field of magnetism, the topological protection inherent to skyrmions ensures their stability even at nanoscales. This unique attribute positions magnetic skyrmions as promising carriers for information storage with the potential to enhance storage density and stability.[16–18]

Analogous to their magnetic counterparts, optical skyrmions have garnered significant attention in recent years.[19] Unlike magnetic skyrmions, optical skyrmions are not confined to near-field surface waves,[20–22] they can also be generated in free space.[23–33] The choice of three-dimensional vector field for constructing optical skyrmions varies widely based on propagation conditions and application scenarios, including electric field vectors,[20,21,34] Poynting vectors,[35] spin angular momentum,[36–38] Stokes vectors.[24,27,28,30,39,40]. Among these, the Stokes vector field is favored as the primary platform for optical skyrmions, owing to the experimental simplicity of measuring Stokes parameters. Under paraxial conditions, Stokes skyrmions are typically constructed by superposing Laguerre-Gaussian beams with distinct orbital angular momenta and orthogonal polarizations.[23,29] While these skyrmions propagate alongside the beam and retain a stable skyrmion number, their large transverse dimensions limit practical applications—such as high-density information storage. In contrast, high numerical aperture (high-NA) conditions offer a promising solution to overcome this limitation: the tightly focused fields produced under high-NA illumination exhibit a drastically reduced transverse scale and more abrupt polarization variations, which significantly enhance the spatial confinement of skyrmions and thereby boost their information density.

Two primary methods exist for constructing skyrmions under high-NA conditions. The first involves generating Stokes skyrmions by superposing and tightly focusing two specialized beams with orthogonal complex amplitude modes and opposite polarizations—typical examples include Laguerre-Gaussian and Bessel-Gaussian beams.[24,28,41] A fundamental limitation of this method, however, lies in the inherent strong longitudinal electric field component induced by high-NA focusing; this component disrupts the purity of the Stokes skyrmion. The second approach entails the tight focusing of an azimuthally polarized optical vortex (APOV), which produces structures with even-order skyrmion numbers.[31,42] Since azimuthally polarized beams exhibit no longitudinal field component after focusing, all energy is fully confined to forming the transverse Stokes skyrmion texture—rendering this approach particularly suitable for constructing high-NA Stokes skyrmions with well-defined topological features. Notably, the polarization phase modulation technique (underpinning the second approach) relies on complex APOV generation systems and is restricted to skyrmions construction only in the focal plane. As skyrmions propagate before and after the focal plane, they degrade rapidly—this also severely impedes their ability to transmit information in free space.

In this paper, we demonstrate the generation of a second-order, non-diffractive optical skyrmion needle with subdiffraction confinement under high-NA conditions, leveraging the intrinsic polarization invariance of traditional optical needles. The resulting Stokes skyrmion field retains its topological features during propagation. To generate this field, we employ a single plasmonic metalens that functions simultaneously as a polarization filter and a super-resolving focusing element. This metalens converts left-handed circularly polarized (LCP) light into an APOV—eliminating the need for bulky conventional APOV-generation systems—and tightly focuses the APOV to form the super-oscillatory skyrmion needle. Notably, a tightly focused APOV supports essentially no longitudinal electric field, ensuring all optical energy is dedicated to constructing a pure transverse Stokes texture. Experiments and simulation results confirm non-diffractive propagation over a 4 μm depth of focus. Throughout this range, the transverse intensity full width at half maximum (FWHM) remains below $0.44\lambda$, while the FWHM of Stokes parameter $S_3$ stays under $0.60\lambda$. This distinctive skyrmion distribution opens promising avenues for applications in optical communication, nanoscale displacement sensing.

## 2. Principle

As mentioned above, generating an APOV beam via polarization-phase modulation followed by tight focusing can produce even-order skyrmions at the focal point, where the skyrmions order $N_{sk}$ is tied to the beam's orbital angular momentum $l\hbar$ ($N_{sk} = 2l$). However, the polarization texture degrades rapidly within the FWHM range along the optical axis, failing to retain the three-dimensional (3D) characteristics of skyrmions. As such, these structures are only describable as 2D skyrmions (or "baby skyrmions"). Given that optical needles exhibit inherent polarization invariance, we propose designing a super-resolved optical needle field to address two key challenges: the inability of skyrmions to propagate in 3D under high-NA conditions, and the need to further reduce skyrmion size. To quantitatively assess the quality of the generated skyrmion field, the skyrmion number ($N_{sk}$) is commonly used as a metric. Physically, this parameter describes how many times the polarization states across the entire optical field cover the sphere of possible Stokes vectors within a fixed region. It is calculated using the following formula:

$$N_{sk} = \frac{1}{4\pi} \iint_{\sigma} \boldsymbol{n} \cdot \left( \frac{\partial \boldsymbol{n}}{\partial x} \times \frac{\partial \boldsymbol{n}}{\partial y} \right) dA \qquad (1)$$

where $\boldsymbol{n} = (S_1, S_2, S_3)/S_0$ represents the spatial distribution of the normalized Stokes vector. Here $S_0$, $S_1$, $S_2$, $S_3$ are the Stokes parameters of the focal field, defined as:

$$S_1 = I_H - I_V$$
$$S_2 = I_D - I_A$$
$$S_3 = I_L - I_R$$
$$S_0 = \sqrt{S_1^2 + S_2^2 + S_3^2} \qquad (2)$$

Here, $I_H, I_V, I_D, I_A, I_L,$ and $I_R$ denote the intensities of the horizontal, vertical, diagonal, anti-diagonal, left-handed circular, and right-handed circular polarization components, respectively.

To realize a tightly focused, polarization-stable optical needle field capable of supporting 3D skyrmions, we employ a metalens designed by combining vectorial angular spectrum method (VASM)[43] with a binary particle swarm optimization (BPSO). In a Cartesian coordinate system, LCP light is mathematically described by the state vector $|\boldsymbol{E_{in}}\rangle = |\sigma = 1, l = 0\rangle = E_0(|\boldsymbol{H}\rangle + i|\boldsymbol{V}\rangle)$, where $\sigma$ and $l$ denote the spin and orbital angular momentum quantum numbers, respectively, and $|\boldsymbol{H}\rangle$ and $|\boldsymbol{V}\rangle$ represent the horizontal and vertical polarization basis states. Transforming to polar coordinates, the optical field's complex amplitude becomes $|\boldsymbol{E_{in}}\rangle = \sum_{\alpha} |\alpha\rangle\langle\alpha|\boldsymbol{E_{in}}\rangle = E_0 e^{i\theta}(|\boldsymbol{\rho}\rangle + i|\boldsymbol{\theta}\rangle)$, where $|\alpha\rangle \in \{|\boldsymbol{\rho}\rangle, |\boldsymbol{\theta}\rangle\}$ with $|\boldsymbol{\rho}\rangle$ and $|\boldsymbol{\theta}\rangle$ as radial and azimuthal polarization states, respectively. Selectively eliminating the radial component yields an APOV, expressed as $|\boldsymbol{E_{trans}}\rangle = |\sigma = 0, l = 1\rangle = iE_0 e^{i\theta}|\boldsymbol{\theta}\rangle$. As illustrated in Figure 1c, polarization filtering of LCP light is achieved via anisotropic nano-antennas on a delicately designed plasmonic metalens. For right-handed circularly polarized (RCP) light, an APOV with topological charge of -1 can be generated analogously. As illustrated in Figure 1b, our target is a needle-like focal field (design parameters: focal length $f = 10$ μm, depth of focus (DOF) = 4 μm, FWHM = 0.4λ). We aim to generate this target field directly through a BPSO-engineered metalens under circularly polarized illumination—eliminating reliance on high-NA objective lenses for focusing. The metalens is equally divided into 100 concentric annular zones with binary amplitude modulation: each zone is either fully transmissive ($t_i = 1$) or opaque ($t_i = 0$). This zoning strategy mimics a Fresnel zone plate, where binary amplitude patterns enable complex amplitude modulation via interference. Through iterative BPSO optimization under specified YZ-plane constraints (transverse domain $\rho < 2$ μm, longitudinal domain $|z - f| < 2$ μm), the generated field gradually converges to the target profile. The optimization ultimately outputs a binary '01' sequence representing the metalens' amplitude distribution. Table 1 presents the hexadecimal-coded amplitude distribution of the optimized annular metalens. The operational principle is detailed in Figure 1a. The LCP light at 800 nm wavelength illuminates the optimized metalens (radius 30 μm, 300 nm ring width, see Figure 1c). Anisotropic nano-antennas on the metalens—supporting electric dipole resonance—perform polarization filtering by selectively removing the radial component $|\boldsymbol{\rho}\rangle$, generating the APOV. Acting as a high-NA system (NA = 0.95), the metalens simultaneously focuses the amplitude-modulated APOV to form the desired focal field with preset characteristics, as shown in Figure 1d. Notably, the metalens exhibits rotational symmetry: reversing the handedness of $|\boldsymbol{E_{in}}\rangle$ allows the formation of APOV with a topological charge of -1 using the same parameters, with only an opposite polarization sign in the focal field. Therefore, subsequent analysis focuses solely on LCP incident light.

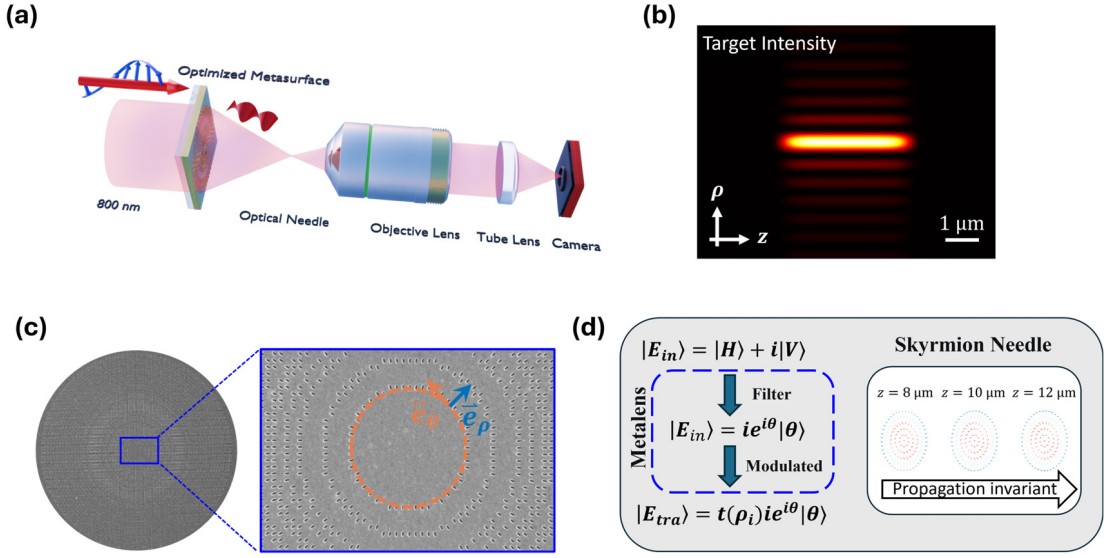

**Figure 1.** (a) Schematic diagram of subdiffraction skyrmion needle generation through the interaction between circularly polarized light and a binary metalens. The arrow indicates circularly polarized incident light, while the spiral structure represents the vortex phase wavefront. (b) Target field distribution within the constrained region ($\rho < 2$ μm, $|z - f| < 2$ μm), serving as the optimization objective. The optical needle is designed with a focal length of 10 μm and DOF of 4 μm. (c) Scanning electron microscope image of a 60 μm-diameter metalens with 100 radially distributed programmable rings (switching states controlled by the optimization result). (d) Generation process of the skyrmion needle, with the inset illustrating its non-diffractive characteristics along the optical needle.

| Table 1. Amplitude Profile of the Optimized Super-oscillatory Metalens $(t_i)$ | |
|---|---|
| No. of rings | Amplitude of ring belt |
| 1-100 | 0 0 8 8 A D A 6 9 B F E 9 A 4 9 2 4 9 2 4 9 2 C 0 |

## 3. Results

### 3.1. Optical Needle

Through rigorous derivation via VASM, the transmitted electric field distribution of an APOV beam $|\boldsymbol{E_{tra}}\rangle = E_0 e^{i\theta}|\boldsymbol{\theta}\rangle$ incident on a binary rotationally symmetric mask can be expressed as:

$$\begin{pmatrix} E_x \\ E_y \\ E_z \end{pmatrix} = \mathcal{F}^{-1} \left\{ \begin{pmatrix} 1 - \sigma_x^2 & -\sigma_x \sigma_y & -\sigma_x \sigma_z \\ -\sigma_x \sigma_y & 1 - \sigma_y^2 & -\sigma_y \sigma_z \\ -\sigma_x \sigma_z & -\sigma_y \sigma_z & 1 - \sigma_z^2 \end{pmatrix} \cdot \mathcal{F} \begin{Bmatrix} -E_0 e^{i\theta} \sin \theta \\ E_0 e^{i\theta} \cos \theta \\ 0 \end{Bmatrix} e^{ik_z z} \right\} \tag{3}$$

where $\boldsymbol{\sigma} \equiv (\sigma_x, \sigma_y, \sigma_z)$ denotes the normalized wavevector, $\mathcal{F}$ and $\mathcal{F}^{-1}$ represent the Fourier transform and inverse Fourier transform, respectively. Figure 2a shows the numerically simulated cylindrically symmetric intensity distribution in the YZ longitudinal plane after light transmits through the metalens, while Figure 2b presents the corresponding experimental measurements—revealing excellent consistency between simulation and experiment. The axial cross-sectional intensity distribution exhibits high uniformity over the range $z = 8-12$ μm, with FWHM of each cross-section remaining below the Abbe diffraction limit ($\lambda/(2NA) = 0.52\lambda = 421$ nm). As shown in Figure 2f, both theoretical and experimental FWHM values of the optical needle's transverse cross-section exhibit subdiffraction characteristics over approximately $5\lambda$. Experimental measurements show focal spot sizes fluctuating slightly between $0.41\lambda$ and $0.44\lambda$. Figures 2c−2d compare the simulated and experimental intensity distributions at the focal plane ($z = 10$ μm), yielding FWHM values of $0.41\lambda$ and $0.42\lambda$ respectively. The radial intensity profiles in Figure 2e exhibit excellent agreement between theory and experiment; minor discrepancies arise from experimental errors during characterization. Notably, as shown in Figure 2e, Notably, as illustrated in Figure 2e, the intensity of the longitudinal component is five orders of magnitude lower than that of the transverse components—confirming the field is essentially pure transverse ($|E_z| \approx 0$). These results verify that, via metalens modulation, we successfully generated a non-diffractive super-resolution optical needle with a 4 μm DOF, featuring FWHM values consistently below $0.44\lambda$ and negligible longitudinal field components. Furthermore, the side-lobes observed near the needle peak are inherent to super-oscillatory optical fields.[44,45]

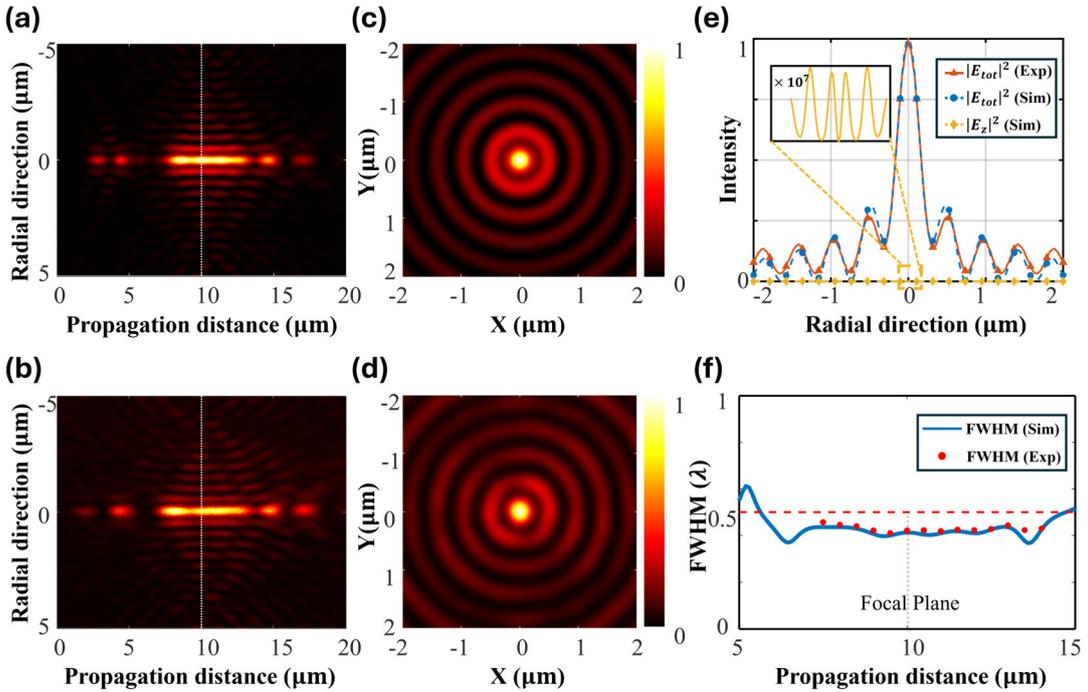

**Figure 2**. (a, b) Numerically simulated (a) and experimentally measured (b) optical intensity distributions over 20 μm propagation distance. (c, d) Focal-plane ($z = 10$ μm) intensity distributions of the generated optical needle: simulation (c) and experimental (d). (e) Horizontal line-scan profiles across the focal spot: simulated total intensity (blue dashed curve with circles) with longitudinal

components (yellow dashed line with diamonds), compared with experimental total intensity (orange solid curve with triangles). (f) Evolution of FWHM along the propagation direction, where the blue solid curve represents theoretical calculation and red solid circles denote experimental data acquired at 0.5 μm intervals from 7.5 μm to 14 μm.

## 3.2. Stokes Skyrmions Needle

Figure 3a presents the simulated results of the Stokes parameters $S_1$, $S_2$, and $S_3$. Notably, $S_1$ and $S_2$ exhibit a rotationally symmetric relationship. For $S_3$, its value evolves gradually with radial distance from the beam center: starting at 1 at the center, decreasing to 0 at intermediate radii, then dropping sharply to –1. The black dashed curve indicates the radial position where $S_3 = -1$ ($\rho = 0.35$ μm). Physically, this $S_3$ evolution corresponds to a clear polarization state transition: the beam center consists of pure LCP light; as radial distance increases, the polarization gradually becomes linear polarization; finally, it transitions rapidly to pure RCP light at the dashed curve. When mapping the real-space distribution onto the Poincaré sphere (polarization parameter space), the beam center corresponds to the sphere's north pole, while the position marked by the black dashed curve aligns with the south pole. As the radial distance $\rho$ increases, the polarization state traces a path from the north pole to the south pole. Critically, for any fixed radial distance $\rho$ between the center and the black dashed curve, a full azimuthal rotation of the beam corresponds to two complete loops around a latitude circle on the Poincaré sphere. This spatial polarization profile is a defining characteristic of a second-order skyrmion structure.

To more clearly illustrate how the normalized Stokes vector covers the Poincaré sphere, we use a hue map for visualization: the hue is determined by $\text{atan}(S_2/S_1)$ which represents the orientation angle of the polarization ellipse; and the lightness is determined by the magnitude of $S_3$, corresponding to the ellipticity. As shown in Figure 3b, the Stokes vector distribution resembles the phase profile of a conventional second-order OAM beam, exhibiting two full azimuthal cycles. This pattern implies double wrapping of the Stokes vector around the Poincaré sphere—consistent with the polarization ellipse distribution in Figure 3e. Experimentally, we measured the six polarization-component intensities (required to calculate the Stokes parameters $S_i$ in Eq. (2)) using the Stokes polarimeter. From these measurements, we derived the corresponding Stokes parameters with results presented in Figure 3b and 3d. Using Eq. (1), we calculated the skyrmion numbers within the black dashed curve: 2.00 for simulations and 1.96 for experiments. These values confirm the successful generation of a second-order skyrmion at the focal plane.

To quantitatively analyze the radial variation of the Stokes parameter $S_3$, Figure 3f plots its radial profile. The blue dashed curves with circles (simulation) and the orange solid curves with triangles (experimental measurement) yield FWHMs of 0.57λ and 0.56λ, respectively. For comparison, the yellow dashed curve with diamonds shows the $S_3$ radial profile of an APOV focused by the same high-NA objective lens (NA = 0.95), as calculated using the Richards-Wolf method;[46] the profile exhibits a larger FWHM of 0.74 λ. This confirms that the metalens-geneereted skyrmion achieves smaller spatial confinement than the traditional diffraction-limit case.

In practical applications, a critial concern arises: if a skyrmion-carrying beam has extremely low light intensity at certain locations while its polarization state traverses the Poincaré sphere, the polarization information at those points may be drowned out by noise and become undetectable. To address this, we calculated the minimum light intensity in the regions where $\rho < 0.35$ μm (Figure 3e). Our results indicate that this minimum intensity is 0.16 times the focal peak intensity, differing by only one order of magnitude. This detectable intensity level mitigates constraints on signal detection and supports the practical utility of these skyrmion structures. We further calculate the polarization inversion width which is defiend as the distance betweeen two adjacent points where $|S_3|$ equals half its maximum value. This width characterizes the spatial scale at which optical skyrmions reverse the polarization state, exhibiting fine subwavelength features ($0.07\lambda$). Collectively, these findings confirm our expectation: we have successfully realized skyrmion fields with subdiffraction confinement in the focal plane.

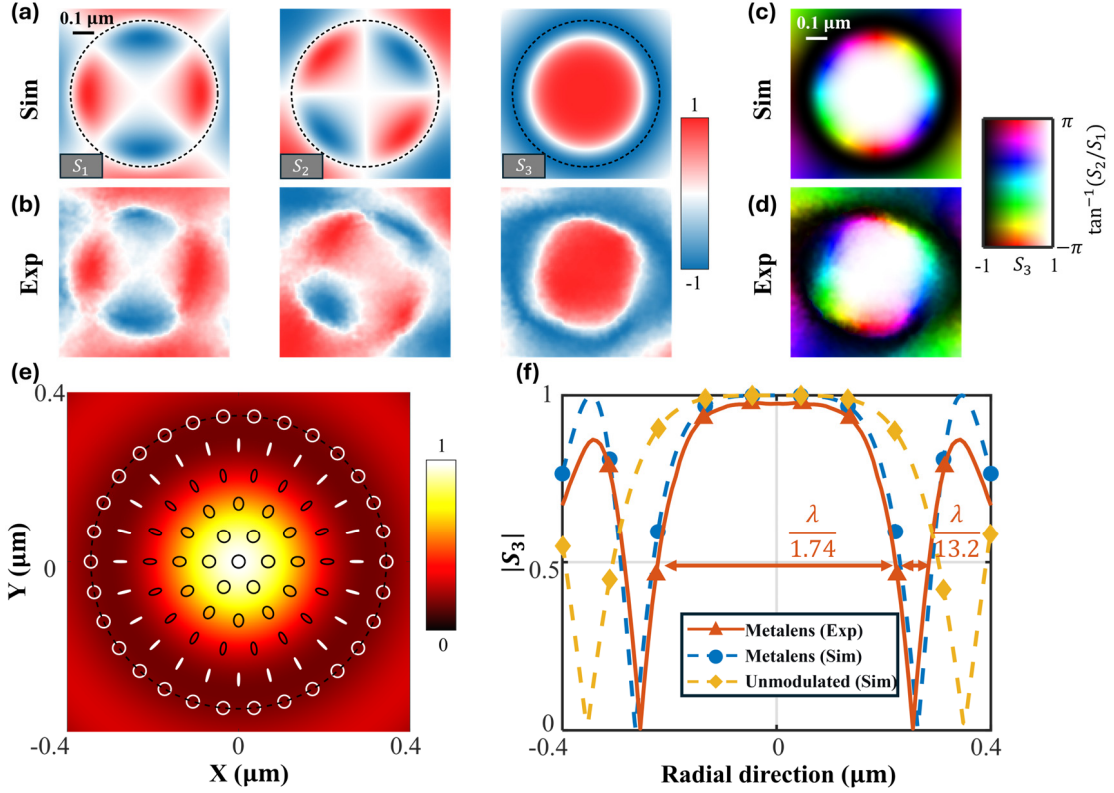

**Figure 3.** (a, b) Distributions of the Stokes parameters $S_1$ at the focal plane ($z = 10$ μm, $\rho < 0.4$ μm): (a) simulated results; (b) experimental measurements. The black dashed curve marks the position where $S_3 = -1$. (c, d) Skyrmion hue maps at the focal plane: white at the center indicates LCP, and black at the edge indicates RCP. The color bar is determined by both the orientation angle and ellipticity of the polarization ellipse, encoding the 3D information of the Stokes vector. (e) Polarization ellipse distribution at the focal plane. The black dashed curve denotes the same feature as in (a). White and black represent left- and right-handed polarization, respectively. (f) Radial profile of the absolute value of the Stokes parameter $|S_3|$: the orange solid curve with triangles denotes experimental results, the blue dashed curve with circles represents simulated values, and the yellow dashed curve with diamonds indicates the theoretical simulation for an APOV focused under the same NA condition without metalens modulation.

Our analysis extends beyond the single focal plane to the skyrmion distribution across the entire optical needle. We numerically simulated and experimentally measured skyrmion distributions at several discrete cross-sections along the optical needle, as presented in Figure 4a and 4b. The Stokes distributions at these cross-sections exhibit no significant differences from those at the focal plane, consistently showing two azimuthal cycles. Notably, no texture transformation—specifically, no rotation of the skyrmion configuration (analogous to the Néel-to-Bloch conversion observed in paraxial systems)—was detected during propagation. This stability arises from the optical needle's intrinsic polarization-invariance. Figure 4c presents the skyrmion number profile along the entire optical needle (blue solid line: simulation; red circles: experiment), revealing that the skyrmion number remains highly stable and exhibits non-diffractive behavior across the full needle length. Simultaneously, the FWHM of $|S_3|$ along the optical needle also stays stable, fluctuating only between $0.55\lambda$ and $0.61\lambda$ (Figure 4d). This constrasts sharply with paraxial systems, where skyrmion profiles expand during propagation from the beam waist to the Rayleigh range. Figure 4d also plots the polarization inversion width along the optical needle; like the skyrmion FWHM, this parameter also exhibits non-diffractive characteristics, varying only between $0.07\lambda$ and $0.08\lambda$. Therefore, these results confirm the non-diffractive propagation of second-order skyrmions: their skyrmion number, texture, size, and polarization inversion width all remain stable. In short, we have achieved a unique non-diffractive and subdiffraction-confined optical skyrmion needle field.

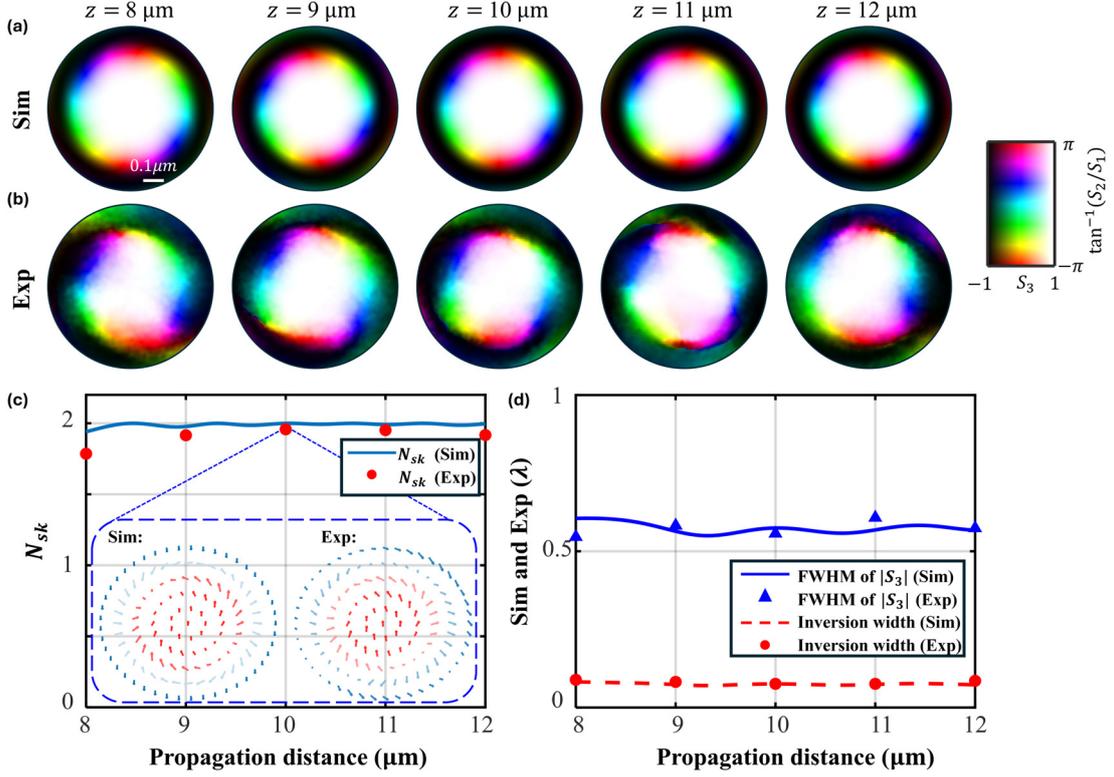

**Figure 4.** (a, b) Hue maps $(\rho < 0.4\,\mu m)$ of the Stokes vector, acquired at $1\,\mu m$ intervals over the axial range $8-12\,\mu m$ beyond the metalens: (a) simulated results; (b) experimental measurements. (c) Evolution of the skyrmion number along the skyrmion needle. The blue solid curve denotes simulated values, and the red solid curve denotes experimental measurements; Shown in the inset are the simulated results and the experimentally measured Stokes vector in the focal plane ($z = 10\,\mu m$). (d) Axial evolution of two metrics derived from $|S_3|$ along the skyrmion needle: the FWHM and the polarization-inversion width. Simulations are shown as blue solid curves and red dash curves; measurements as blue triangles and red circles.

## Conclusions

In summary, we have designed and demonstrated a second-order skyrmion needle that achieves both subdiffraction confinement and non-diffractive propagation under a high numerical aperture (NA = 0.95). The field is generated by a single metalens that integrates two core functions: polarization filtering and subdiffraction focusing. The design leverages two key attributes: the polarization invariance of conventional optical needles and super-oscillatory localization. Unlike previous high-NA approaches for skyrmions generation, this skyrmion structure retains its topological features along the entire needle length and maintains a fixed configuration—free from transformation. Critically, the absence of the longitudinal electric field ensures all optical energy is concentrated in forming a purely transverse Stokes-vector texture. Comprehensive simulations and experiments confirm the non-diffractive propagation of the skyrmion needle. Key parameters remain both stable and sub-diffractive throughout propagation: intensity FWHM, skyrmion number, Stokes parameter FWHM, and polarization inversion width.

This spatially uniform, subwavelength-scale skyrmion distribution offers a novel topological information carrier with ultra-compact dimensions—an attribute particularly advantageous for optical information transfer, where stable transmission and high fidelity are critical requirements. Its subwavelength polarization inversion width also opens a promising avenue for advancing high-precision optical metrology and sensing technologies. Furthermore, the development of single metalens capable of generating skyrmions—with the growing trend toward integrated photonic devices and fiber-based lab-on-a-chip systems, greatly expands the application prospects for skyrmions in future optical technologies.

**Acknowledgement**

This work was supported by the Chinese Academy of Sciences (CAS) Project for Young Scientists in Basic Research (No. YSBR-049) and the Overseas Excellent Youth Science Foundation. Part of this study was conducted at the USTC Center for Micro- and Nanoscale Research and Fabrication.